\documentclass[reprint,amsmath,amssymb,aps,prl,superscriptaddress]{revtex4-1}

\usepackage{times,color,amsmath}
\usepackage{subfigure,epsfig,graphicx,epstopdf}
\usepackage[colorlinks]{hyperref}
\hypersetup{linkcolor = {blue},	citecolor = {blue},	urlcolor = {blue}}

\newcommand{\affa}{\affiliation{Department of Applied Physics, KTH Royal Institute of Technology, Albanova University Centre, Roslagstullsbacken 21, 106 91 Stockholm, Sweden}}
\newcommand{\affb}{\affiliation{Nordita, Stockholm University and KTH Royal Institute of Technology, Hannes Alfvéns väg 12, SE-106 91 Stockholm, Sweden}}
\newcommand{\affc}{\affiliation{Institute for Physics of Microstructures, Russian Academy of Sciences, 603950 Nizhny Novgorod, GSP-105, Russia}}
\newcommand{\affd}{\affiliation{Center for Integrated Quantum Information Technologies (IQIT), School of Physics and Astronomy and State Key Laboratory of Advanced Optical Communication Systems and Networks, Shanghai Jiao Tong University, Shanghai 200240, China}}
\newcommand{\affe}{\affiliation{Department of Physics, University of Connecticut, Storrs, Connecticut 06269, USA}}

\begin{document}

\title{Coexistence of extended and localized states in finite-sized mosaic Wannier-Stark lattices}

\author{Jun Gao} \email{junga@kth.se} \affa
\author{Ivan M. Khaymovich} \email{ivan.khaymovich@gmail.com} \affb\affc
\author{Adrian Iovan} \affa 
\author{Xiao-Wei Wang} \affd
\author{Govind Krishna} \affa 
\author{Ze-Sheng Xu} \affa	
\author{Emrah Tortumlu} \affa
\author{Alexander V. Balatsky} \affb \affe
\author{Val Zwiller} \affa 	
\author{Ali W. Elshaari} \email{elshaari@kth.se} \affa 

\date{\today}

\begin{abstract}

Quantum transport and localization are fundamental concepts in condensed matter physics. It is commonly believed that in one-dimensional systems, the existence of mobility edges is highly dependent on disorder. Recently, there has been a debate over the existence of an exact mobility edge in a modulated mosaic model without quenched disorder, the so-called mosaic Wannier-Stark lattice. Here, we experimentally implement such disorder-free mosaic photonic lattices using silicon photonics platform. By creating a synthetic electric ﬁeld, we could observe energy-dependent coexistence of both extended and localized states in \textbf{a finite number} of waveguides. The Wannier-Stark ladder emerges when the resulting potential is strong enough, and can be directly probed by exciting different spatial modes of the lattice. Our studies provide the experimental proof of coexisting sets of strongly localized and conducting (though weakly localized) states in finite-sized mosaic Wannier-Stark lattices, which hold the potential to encode high-dimensional quantum resources with compact and robust structures.

\end{abstract}

\maketitle

The phenomenon of localization of electronic Bloch waves was first studied by P. W. Anderson~\cite{Anderson1958}, where electronic wave functions become exponentially localized due to disorder. For a 3D system, there exists a mobility edge (ME), which separates the localized and extended states by a critical energy as a function of disorder level~\cite{Evers2008}. Lower dimensional models, by replacing disorder with a quasiperiodic potential, for example, the Aubry-Andr{\'e} Harper model, can host both localized and extended states, showing an energy-independent critical transition at a self-dual point~\cite{AAH1,AAH2,AAH3}. By further incorporating long-range hopping~\cite{Prange1983,Biddle2009,Biddle2010,Gopalakrishnan2017,Deng2019,Saha2019,Deng2018,Nosov2019,Kutlin2020,Deng2022}, varying on-site potential~\cite{DasSarma1988,DasSarma1990,Ganeshan2015,Liu2022}, breaking self duality~\cite{Li2017,Yao2019,Yin2020}, applying periodic drive~\cite{Roy2018,Ray2018,Sarkar2021,Gonsavles2023}, considering flat-band systems~\cite{Danieli2015,Ahmed2022,Lee2022,Kim2022}, or introducing a quasiperiodic potential in mosaic lattices~\cite{Wang2020,Liu2021}, the modified model could support an energy-dependent ME in the energy spectrum. So far, the existence of ME in low dimensional systems has been experimentally confirmed with ultra-cold atomic lattices~\cite{Ucold1,Ucold2,Ucold3}, it is natural to ask if random or quasiperiodic potential is essential for a system to manifest MEs.

A recent study~\cite{Dwiputra2022} claimed that a disorder-free 1D mosaic lattice with Stark effect could exhibit an exact ME, where such disorder-free localized states can be tracked back to the famous Wannier-Stark lattice~\cite{Wannier1962,Fukuyama1973,Emin1987}. By introducing a static electric field to the lattice, the resulting potential may lead to exponential localization of the wave function. With strong enough fields, the Wannier-Stark ladder is recovered, and each energy level corresponds to a localized eigenstate, while all the survived extended states live at small enough energies. However, the theory is exciting but also puzzling, as more recently pointed out by a work of S. Longhi~\cite{Longhi2023}, that Avila's global theory cannot be applied here and Lyapunov exponents cannot be defined for Stark potentials, going to infinity. In the thermodynamic limit (when the system size goes to infinity), all states become localized with the exception of few isolated extended states, thus strictly speaking no disorder-free ME exists. Only under a finite-height mosaic potential~\cite{Wei2023}, can the Wannier-Stark lattice manifest a pseudo ME, which can be experimentally realized and probed.

\begin{figure*}[!t]
	\centering
	\includegraphics[width=0.95\linewidth]{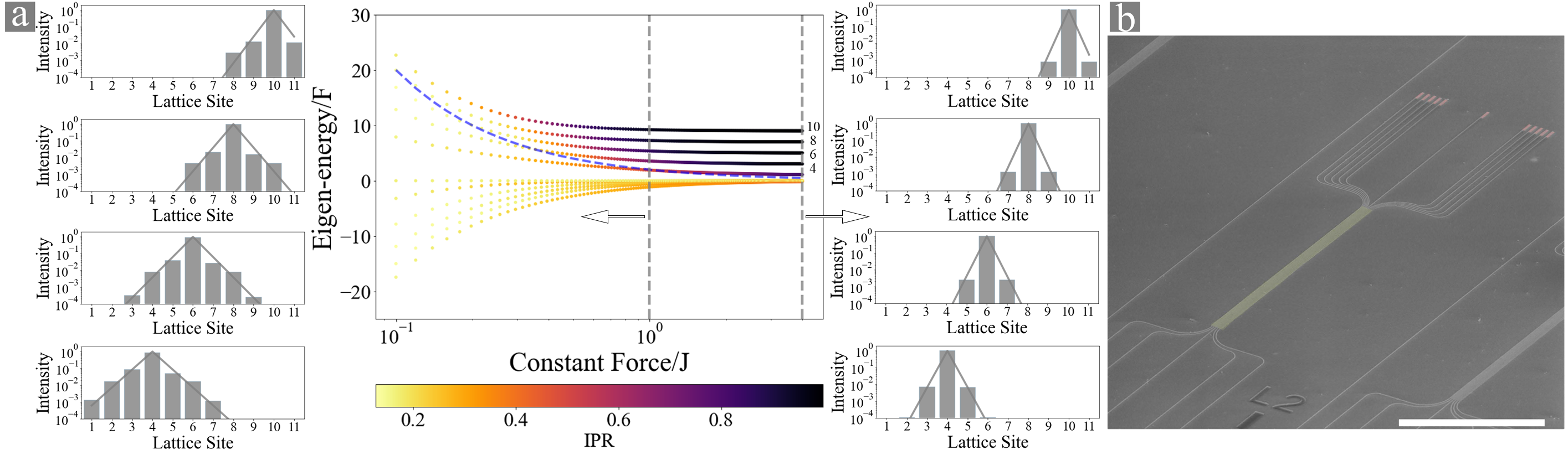}
	\caption{\textbf{Localization of the Wannier-Stark ladder states in disorder-free $\kappa=2$ finite-sized mosaic lattice in a synthetic electric field.} (a)~Eigenenergy spectrum of the photonic mosaic lattice under different constant forces. The force is implemented by augmenting the potential of the even lattice sites through Eq.~\eqref{eq:2}. The lattice comprises of $11$ sites, with a constant hopping parameter $J=0.01$ among the sites. Histograms show the intensity distributions of the eigenstates, localized slightly stronger than exponentially at even sites, in logarithmic scale for (left)~moderate $F/J=1$ and (right)~strong $F/J=4$ force regimes. Lines show exponential fits. The states are indicated in the middle energy diagram by their lattice site numbers. The strong field $F/J=4$ is selected to detect the Wannier-Stark ladder, as at this point, the eigenstates corresponding to high-energy levels in the system are localized, showing equidistant spacing in the energy diagram. In contrast, all the states below the finite-size ME (indicated by a dashed blue line) are extended as light traverses the lattice. (b)~A scanning electron microscope (SEM) image of the nanophotonic device is presented, depicting the photonic lattice, denoted in light green, which features individual lattice sites terminated with a grating coupler, accentuated in red, interconnected through a fan-out structure. The scale bar shown in white has a length of $100$~\textmu m. }
  	\label{fig1}
\end{figure*}

In this Letter, we experimentally realize such finite-sized disorder-free mosaic Wannier-Stark model, observe energy-dependent coexistence of localized and extended states, and probe the Wannier-Stark ladder in the photonic mosaic lattices. By utilizing a $\text{Si}_3\text{N}_4$ waveguide array with engineered on-site potentials and nearest-neighbor hopping terms~\cite{TPRev1,TPRev2,TPRev3,TPRev4}, we create a synthetic constant electric field that localizes part of the eigenstates in real space. The light intensity distribution is directly probed through a single-shot top imaging with a fan-out structure and grating couplers. To verify the coexistence of extended and localized states in the system, we excite the lattice in both moderate and strong force regime, and compare the inverse participation ratio (IPR) of spreading wave packets. In the strongly localized regime, we can directly excite a single mode in the lattice to probe the localization eigenstates, with over 98$\%$ fidelities between experimentally measured states and the corresponding eigenstates. By calculating the overlap weights with respect to the localized eigenstates, we reconstruct the Wannier-Stark ladder in the strong field regime, demonstrating equally spaced energy levels as predicted. Our work extends the understanding of the Wannier-Stark localization and ME physics, in addition to providing a new platform for studying the extended-localization transition, which offers a more scalable and precise approach to modulate the lattice parameters at room temperature. Our results offer a promising avenue for on-chip high-dimensional quantum information encoding~\cite{InteP,Elshaari2020,Chang2023,Zhang2021,Onchipqudit}, and will also inspire particle statistics induced quantum correlation in ME studies by using multi-photon state excitation~\cite{Lahini2010,Segev2013,Crespi2013}.

Here, we consider a 1D mosaic lattice with Stark effect, which can be described by the following Hamiltonian
\begin{equation}\label{eq:1}
H=-J \sum_n\left(c_n^{\dagger} c_{n+1}+\text { H.c. }\right)+\sum_n \epsilon_n c_n^{\dagger} c_n,
\end{equation}
\begin{equation} \label{eq:2}
\epsilon_n= \begin{cases}F n, & n= m\kappa\text{, with integer~}m, 
\\ 0, & \text { otherwise, }\end{cases}
\end{equation}
where $c_{n}$ is the annihilation operator at site $n$, $J$ is the nearest-neighbor hopping term, $\epsilon_n$ is the on-site modulated potential, which is further determined by a constant force $F$ and modulation period $\kappa$. This model introduces a Stark potential on every $\kappa$th site, and the value is linearly increasing along the site number. Note that in the experiment, the total number of lattice sites is always limited, which sets an upper bound for the applied potential.

By studying separately the localized wave functions, having most of their weight at $n=m\kappa$ sites, decaying faster than exponentially with distance, and large energies $E_m \simeq F m \kappa \gg 1$ with the perturbation theory in $J/E_m$ and the corresponding effective Hamiltonian with the projected out above localized states~\cite{SM}, we have improved the results of~\cite{Dwiputra2022}, finding the following pseudo ME (the blue line in the middle energy diagram in Fig.~\ref{fig1}(a)) in a finite-size system at $\kappa > 1$
\begin{equation}\label{eq:3}
E = E_{ME} = \max\left[2J, \left(\frac{e J}{F \kappa L}\right)^{1/(\kappa-1)}\right] \ .
\end{equation}
In~\cite{SM} we have also provided $\kappa-1$ exact plane-wave states with the momenta $p=q\pi L/\kappa$, $q=1,\ldots, \kappa-1$ and antinodes at $n=m\kappa$, as well as shown that the rest extended states have reduced values at $n=m\kappa$. The plane-wave states are unaffected by the potential and form the subband of the width $2J$, with energies $E_q = -2J \cos(\pi q/\kappa)$. All this is in agreement with the analytical solution given in~\cite{Longhi2023} for the thermodynamic limit.

Equation~\eqref{eq:3} implies that in the strong field regime or large system size, the pseudo ME will converge to $2J$ (at experimentally available system sizes), and all extended states share the property of small eigenstate coefficients at the sites $n=m\kappa$ with the potential~\cite{SM}, while the localized eigenstates characterized by the exponential decay and IPR form a Wannier-Stark ladder, as shown in Fig.~\ref{fig1}(a).

To experimentally realize the mosaic lattice, we first provide elaboration on the design of the photonic lattice required to simulate the synthetic static electric field, which corresponds to a linearly increasing on-site potential. We use full-vectorial mode solver~\cite{silicon,lumerical} to numerically simulate the propagation constants of the waveguide. To illustrate this in~\cite{SM}, we show numerical simulations of the on-site energy for a single isolated waveguide with a fixed height of $250$~nm and varying the width, all relative to the energy at a width of $500$~nm. The fundamental transverse electric (TE) mode is used in all the calculations. Our chosen operating point enables a significant tuning range around the $500$~nm width, while also ensuring single-mode operation. We then express the on-site potential $V$ as a third order polynomial expansion of the waveguide width as follows 
\begin{equation}\label{eq:4}
V(x)=\sum_{m=0}^{3} a_mx^m,
\end{equation}
which helps us to back calculate the waveguide width at the given modulation amplitudes.

\begin{figure}[!t]
	\centering
	\includegraphics[width=0.95\linewidth]{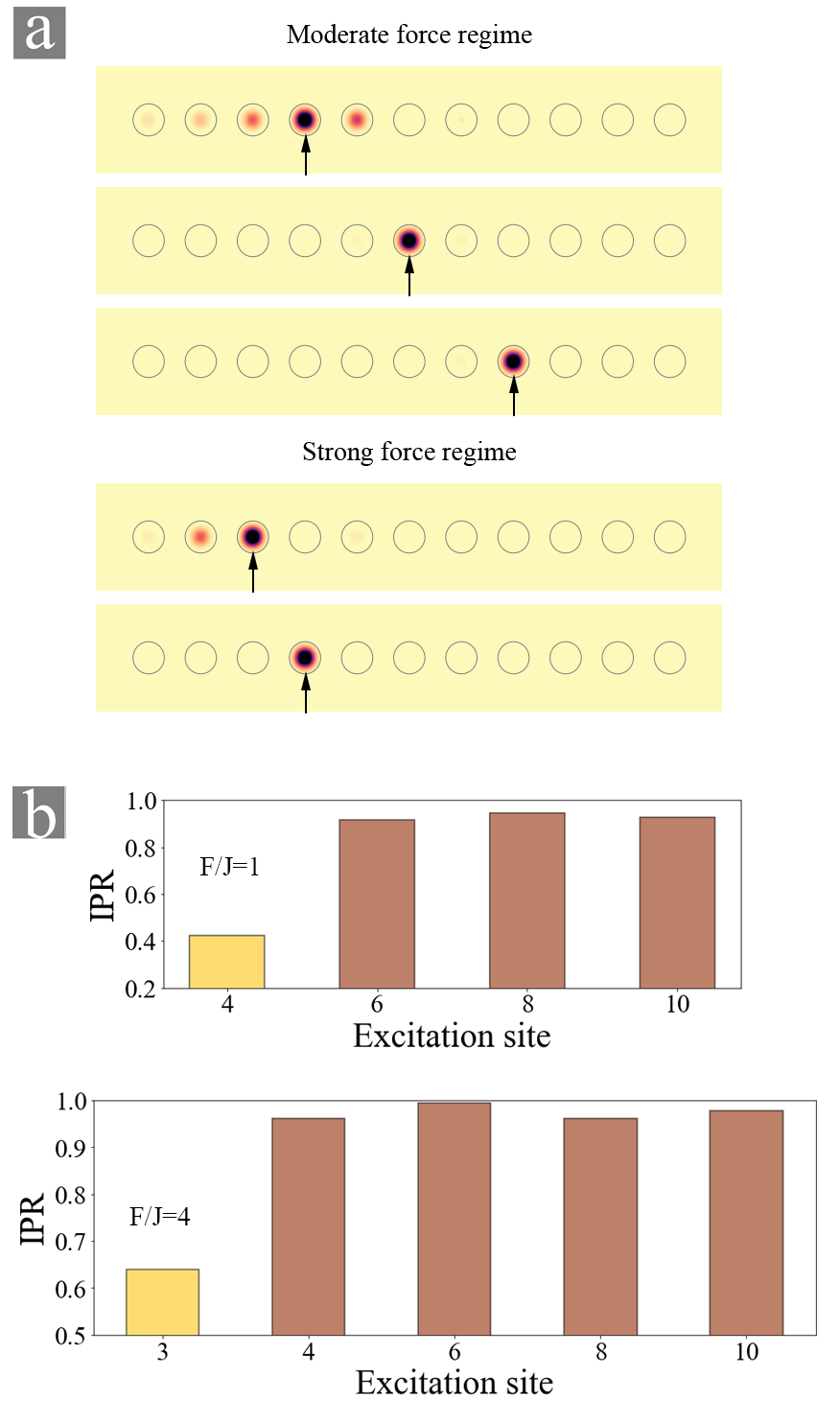}
	\caption{\textbf{Moderate- and strong-field limits and IPR of single-site excitations in the lattice.}
(a)~Output intensity distribution comparison for the single-site excitation at even ($4$th, $6$th, and $8$th) and $3$rd waveguides, respectively. In the moderate-field regime (at the propagation length of $600$~\textmu m), $F/J=1$, the low-energy even-site excitation ($4$th) shows extended feature, while the rest of high-energy states are still strongly localized. At strong fields (at the propagation length of $800$~\textmu m ), $F/J=4$ (bottom), even waveguide ($4$th) excitation results in the localization of light, while the odd ($3$rd) waveguide excitation leads to the spread of light to neighboring lattice sites.  
(b)~The wave-packet inverse participation ratio (IPR), extracted from the data in (a), for various single-site excitation cases at $F/J=1$ and $F/J=4$.}
	\label{fig2}
\end{figure}

In our experimental design, we have chosen a total number of $11$ waveguides, $\kappa=2$, a constant force of $F/J=1$ and $4$ along with a nearest-neighbor hopping term of $J=0.01$. This particular combination has allowed us to operate in both the moderate and strong field limit of Fig.~\ref{fig1}(a), thereby enabling clear identification of both the finite-size extended states and the Wannier-Stark ladder. In the very weak regime, the model is simplified to the quantum-walk framework, which has been experimentally explored~\cite{Jiao2021}. For the lattice design, we solve Eq.~\eqref{eq:4} to retrieve the widths of the modulated waveguides, then engineer the on-site energy to increase it linearly every second waveguide, as expressed by Eq.~\eqref{eq:2}.

After determining the waveguide widths that render the on-site potential throughout the lattice, it is important to mention that in order to maintain a constant hopping term of $J=0.01$ while modulating the on-site energy, the gaps between the waveguides must be carefully designed. A dedicated section in~\cite{SM} is reserved for discussing the employed coupled mode theory and the mode-solver procedure for asymmetric coupled waveguides. It also provides detailed information on the device parameters used for the nano-fabrication. Another section focuses on the device fabrication and the proximity correction of the electron-beam lithography to realize the dense structure of the photonic lattice~\cite{Gao2022,Xu2022}. Fig.~\ref{fig1}(b) shows a scanning electron microscope (SEM) image of the fabricated device, it comprises a photonic mosaic lattice (highlighted in light green), in conjunction with a fan-out configuration of all the lattice sites. The fan-out culminates in grating couplers (highlighted in light red) that facilitate top imaging for measuring the intensity in each lattice site. To examine the excitation dynamics in the lattice, devices with different propagation lengths are nano-fabricated in intervals of $200$~\textmu m. In order to avoid any cross-talk between the excitation sites, two uniform adjacent lattices are employed for each length, with one tailored to excite the even sites and the other to excite the odd sites.

\begin{figure*}[!t]
	\centering
	\includegraphics[width=0.95\linewidth]{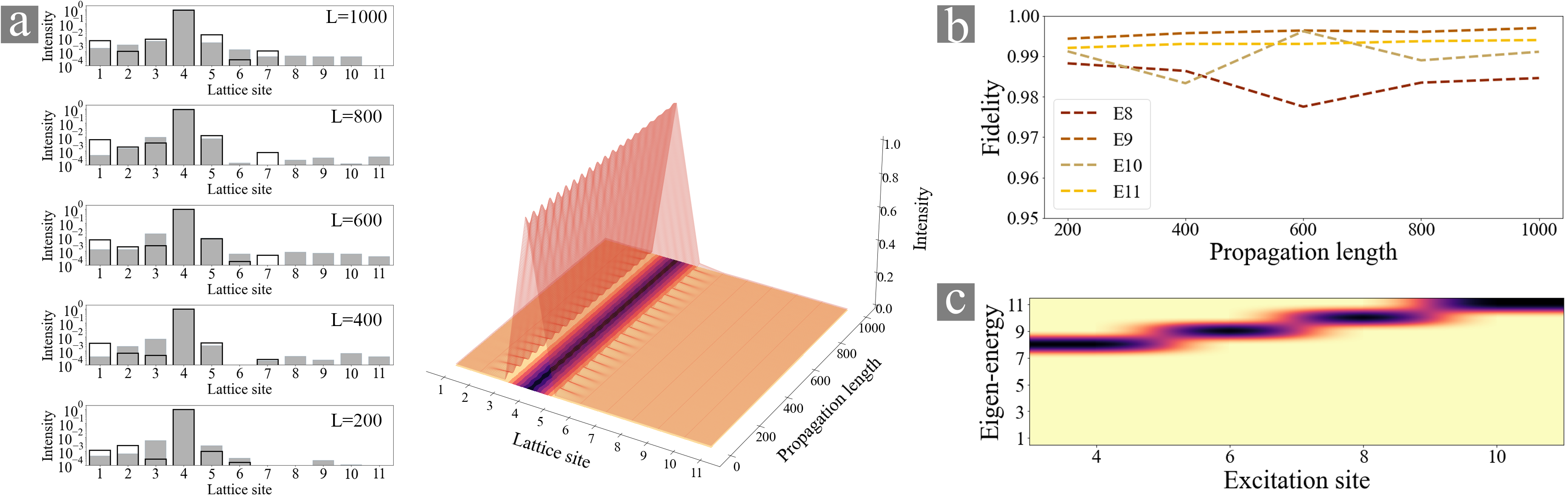}
	\caption{
 \textbf{Localization of the Wannier-Stark ladder states, their fidelity at $F/J=4$, and reconstruction of Wannier-Stark ladder.} {(a)~Numerical simulation of the $4$th waveguide's excitation in the lattice is showcased in the $3$D plot, which demonstrates a strong overlap with the first localized eigenstate of the Wannier-Stark ladder in Fig.~\ref{fig1}.
 Experimentally measured (numerically calculated) intensity distributions in log scale across all $11$ lattice sites, measured at corresponding intervals of $200$~\textmu m, shown by grey (empty) bars. We observe an exceptional agreement between the simulated lattice and the experimental measurements, where the light is primarily confined to the $4$th waveguide with a profound overlap with the localized Wannier-Stark ladder eigenstate.}
 (b)~Fidelity of the even waveguide excitations to the Wannier-Stark ladder eigenstates located above the pseudo ME, extracted from the experimental intensities. The high values of fidelity obtained from the experimental measurements indicate a strong overlap between the single-site excitations in the fabricated lattice and the corresponding theoretical intensity distributions of the eigenstates. To investigate the propagation dynamics of the excitations, fidelity is evaluated at various lengths with a step size of $200$~\textmu m.
 (c)~The calculated weights over different eigenenergies from the experimental data, revealing the Wannier-Stark ladder structure with equally spaced energy levels.}
	\label{fig3}
\end{figure*}

Coherent laser centered at $786$~nm is employed for the excitation of the photonic mosaic lattice. The light is coupled to the nanophotonic chip via a lensed fiber attached to a $6$-axis nano-positioning stage. The TE mode of the waveguide is selectively excited using a fiber-based $3$-paddle polarization-controller at the input. The output light intensity at each lattice site is captured by the grating coupler termination using top imaging with a $40$X objective, and recorded using a charge-coupled device (CCD) camera. Further information on the design and numerical simulations of the fabricated grating couplers, in addition to the top images acquisition and analysis routine, can be found in~\cite{SM}.

To verify the coexistence of localized and extended states in the system, we probe both moderate and strong force regimes by exciting different sites in the photonic lattices. In the moderate regime, there are actually two kinds of extended states. As shown in Fig.~\ref{fig1}(a), there always exists extended states below the pseudo ME, and part of the states above the pseudo ME show extended features due to a weaker modulation amplitude. The upper panel of Fig.~\ref{fig2}(a) demonstrates one example of excitations above the pseudo ME for the moderate force regime. Specifically, we present the measured light distribution after $600$~\textmu m propagation length, for even ($4$th, $6$th, $8$th) sites. With the increasing potential amplitude along the lattice sites, the higher energy states ($6$th and $8$th) clearly show stronger confinement in the excitation site compared to the $4$th site. As a comparison in the strong field regime, the $4$th site is already highly confined, and only the odd site excitation ($3$rd in the lower panel of Fig.~\ref{fig2}(a)) shows the extended feature, which corresponds to contribution from states below the pseudo ME. In order to further explore the localization properties of the lattice, we have computed the IPR, which is a well-established measure in the field of condensed matter physics~\cite{Thouless1974,Borgonovi2016}, $\mathrm{IPR}=\sum_{j} |\psi(j,t)|^4$ of the propagated single-site excitation $\psi(j,t)$. In particular, it provides a quantitative measure of the extent to which an eigenstate is spread out over the entire system or is confined to a small region~\cite{SM}. The resulting IPR values are displayed in Fig.~\ref{fig2}(b), where we observe that high-energy excitations at even lattice sites approach the limit of a single-site localization (IPR$=1$), indicating the highly confined nature of these states. In contrast, the $4$th (moderate regime) and $3$rd (strong regime) waveguide excitation exhibits lower IPR values, indicating a more extended character of this state. The distinct difference between the moderate and strong field regime clearly shows a strong evidence for the existence of energy-dependent ME in the finite-size photonic mosaic lattice with Stark effect.

Then, we focus on the strong force regime to probe the Wannier-Stark ladder. Experimental measurements presented in Fig.~\ref{fig3}(a) exhibit the excitation of the $4$th waveguide in the lattice, corresponding to the first localized state in the ladder of Fig.~\ref{fig1} above the finite-size ME in the strong field regime. The intensity distribution of light across all lattice sites is monitored at regular intervals of $200$~\textmu m. Evidently, the majority of the light intensity is concentrated within the $4$th site, in agreement with the theoretical predictions of the Wannier-Stark ladder. The model predicts the existence of eigenstates that are spatially localized as light propagates within the lattice. Numerical simulation of light dynamics, displayed in the $3$D plot, is conducted utilizing the experimental design parameters, whereby a single-site excitation with considerable overlap with the Wannier-Stark ladder's eigenstates reveals the expected spatial localization.

Similarly, we measure all the other localized eigenstates above the pseudo ME with single-site excitation. In order to assess the quality of the experimental results, we evaluate the fidelity between the measured output state intensity distribution $I^{exp}$ and the corresponding theoretical ladder states $I^{eigen}$, given by the following formula
\begin{equation}\label{eq:5}
Fidelity = \frac{\sqrt{\sum_{n} I_{n}^{exp}\cdot I_{n}^{eigen}}}{\sum_{n}I_{n}^{exp}\sum_{n}I_{n}^{eigen}}.
\end{equation}
We compare the fidelity of each excitation at different propagation distances $z$. Fig.~\ref{fig3}(b) shows the fidelity of the measured four Wannier-Stark ladder states beyond the ME, as a function of the propagation length. At a strong field regime characterized by $F/J=4$, the measured fidelity surpasses $98\%$. This finding is significant as it demonstrates the reliability of the theoretical models used to describe the system under investigation. Additionally, the high fidelity indicates that a single-site excitation is expected to yield an excellent agreement between the output state distribution measured experimentally and the corresponding theoretical Wannier-Stark ladder states.

With the measured localized state intensity distributions (for example, at the propagation distance of $800$~\textmu m), we can reconstruct the Wannier-Stark ladder in the energy spectrum. To achieve this, we calculate the overlap weight of different eigenenergies by projecting the output distribution over all $11$~eigenstates. Note that the measured intensity is strongly localized, thus we can focus on the localized site and ignore the phase information of the other lattice sites. The weight $\{w_{i}\}$ is given by $w_{i}=|\langle \phi_{i}|\sqrt{I^{exp}}\rangle|^{2}$, here $|\phi_{i}\rangle$ represents the $i$th eigenstate. As shown in Fig.~\ref{fig3}(c), we list all the weights in a color map for four Wannier-Stark ladder states. The result clearly reveals the dominant overlap in each energy level, and forms a ladder structure with equally spaced energy levels. We can directly map the Wannier-Stark ladder to different spatially localized modes.

In conclusion, our experimental investigation of the Wannier-Stark ladder and ME in a finite-sized disorder-free photonic lattice has provided new insights into the fundamental concepts of quantum transport and localization in condensed matter physics. Our use of a $\text{Si}_3\text{N}_4$ waveguide array with an engineered on-site potential and nearest-neighbor hopping rate allows us to create a synthetic electric field that localizes part of the eigenstates in real space. By probing the light intensity at each lattice site through a single-shot approach using a fan-out structure and grating couplers, we are able to observe the coexistence of both extended and localized states in the system. Our results offer experimental evidence of recent theoretical works, and have demonstrated the emergence of the Wannier-Stark ladder when the electric field is strong enough. The potential applications of our developed photonic devices are vast and promising, as they offer a compact and robust means of encoding high dimensional quantum resources, for example, the structure can be used to encode dual-rail qubits or even qudits~\cite{moody20222022,gao2023scalable, Chang2023,elshaari2021deterministic, elshaari2017chip}.

\begin{acknowledgments}

The authors would like to thank Prof. Stefano Longhi for helpful discussions and suggestions. J.G. acknowledges support by Natural Science Foundation of Shanghai (Grant No. 20ZR1426400), A.W.E acknowledges support Knut and Alice Wallenberg (KAW) Foundation through the Wallenberg Centre for Quantum Technology (WACQT), Swedish Research Council (VR) Starting Grant (Ref: 2016-03905), and Vinnova quantum kick-start project 2021. V.Z. acknowledges support from the KAW and VR.
Work at Nordita was supported by European Research Council under the European Union Seventh Framework ERS-2018-SYG HERO, KAW 2019.0068 and the University of Connecticut.

\end{acknowledgments}

\end{document}